\documentclass[a4paper,12pt]{iopart}
\usepackage{amssymb}
\usepackage[dvips]{graphicx}
\begin{document}
\title[A.Novikov-Borodin. Off-site continuums \& methods of math. description]{Off-site continuums and methods of their mathematical description} 
\author{A.Novikov-Borodin\dag\ }
\address{\dag\ Institute for Nuclear Research of RAS,\\ 60-th October Anniv.pr. 7a, 117312 Moscow, Russia.}
\ead{novikov@inr.ru}
%\today
\begin{footnotesize} \hspace{100mm} \today \end{footnotesize} 
%
% insert abstract here
\begin{abstract} 

The hypothesis concerning the off-site continuum existence is investigated from the point of view of the mathematical theory of sets. The principles and methods of the mathematical description of the physical objects from different off-site continuums are introduced and discussed. There are also proposed the mathematical methods of the description of the off-site continuum observable structures and the visual perception of its physical objects from the continuum of the observer.  

\end{abstract}
%
%
%\submitto{}
%
\pacs{04.20.Cv; 12.90.+b; 03.65.Bz; 04.50.+h}
%\pacs{ \\04.20.Cv -- Fundamental problems and general formalism in general %relativity; \\12.90.+b -- Miscellaneous theoretical ideas and models of elementary %particles;\\03.65.Bz -- quantum mechanics, foundations; \\04.50.+h -- alternative %theories of gravity. } 
%\\11.30.Cp -- Lorentz and Poincar%\'e invariance in particles and fields;03.30.+p -- special relativity;  
%\tableofcontents
%\maketitle
%\frenchspacing
% body of paper here
\section*{\bf Introduction}

The general theory of relativity generates the conservation laws inside itself  -- and not as a consequence of field equations, but as identities (E.Schr\"odinger \cite{Schr86}). Four identical ratios between Hamiltonian derivatives of some invariant density $ {\mathfrak {R}} $ may be obtained only from one fact of the general invariance of the Integral: 

\begin{equation}
{\cal I} = \int_{\it G} \mathfrak{R} \, d^{4} x ,
\label{INT}
\end{equation}  

and these ratios seem like conservation laws. The continuum $ {\it G} $ is a four-dimensional space-time corresponding to our ``material world''. Physics exactly deals with the description of this continuum: physical laws and objects in it. It seems logical, but do we have some restrictions to the existence of another, off-site continuums, differ from $ {\it G} $? Is our continuum really the only one? We will try to investigate these questions in this paper and will try to find some methods of the mathematical description of such continuums, if they may exist. 

\section{\bf Space-time continuums}
\label{sec:Cont}

Let's suppose, that there exist some continuums $ \tilde{\it G} $ differed from $G$. We will call such continuums as {\it off-site} ones. From the mathematical theory of sets point of view there are a lot of possibilities for such continuums: $ \tilde{\it G} \cap G = $ { \O}; $ \tilde{\it G} \cap G \neq $ { \O}; $ \tilde{\it G} \supset G $; $ \tilde{\it G} \subset G$, etc. If the off-site continuum $ { \tilde G} $ can be observed completely in area $ D $ from $ {\it G} $: $ \tilde {G} \rightleftharpoons D \subset {\it G} $, we shall name $ \tilde G$ as an {\it enclosed} continuum in relation to $ {\it G} $, and $ {\it G} $ will be called as a {\it containing} one in relation to $\tilde G$. 

\begin{figure}[ht]
	\begin{center}
	\includegraphics*[height=85mm, angle=270]{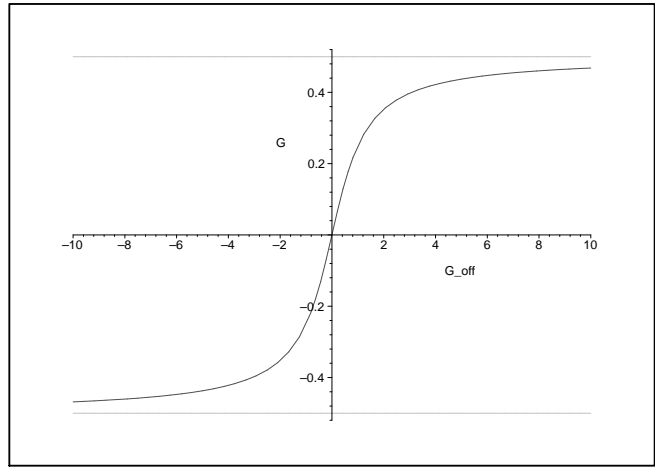}
	\caption{Example of transformations between continuums.}
	\label{fig:Atan}
	\end{center}
\end{figure}

To give an elementary mathematical example of off-site continuums, let's consider two identical $n$-dimensional continuums:  $ {\it G} $: $ \{ x: (x^0, x^1, \dots , x^n) \} $ and $\tilde{\it G} $: $ \{ \tilde{x}: (\tilde{x}^0, \tilde{x}^1, \dots , \tilde{x}^n) \} $ connected with each other by functional coordinate correspondences. The coordinate correspondences $ x^i = (1/\pi) \arctan (\tilde{x}^i) $, $i=0..n $ reflect an $n$-dimensional continuum $ \tilde{\it G} $ inside an $n$-dimensional unit cube from $ {\it G} $: $\tilde{\it G} \rightleftharpoons D_1 \subset {\it G}$. This type of transformations is shown on \Fref{fig:Atan}. Another transformations $\tilde{x}^i = (1/\pi) \arctan(x^i) $, $i=0..n$ reflect $ {\it G} $ inside an $n$-dimensional unit cube from $\tilde{\it G } $: ${\it G} \rightleftharpoons \tilde{D}_1 \subset \tilde {\it G}$. 

All of these continuums have the same {\it power of set}, so, both $ G $ and $\tilde{\it G} $ may be chosen as a continuum of the observer. Generally, it is asserted in the mathematical theory of sets that any multi-dimensional continuum have the same power as a unit line segment [0,1]. So, for example, mathematically, one-dimensional continuum is identical to two-, three- or any $n$-dimensional one.  

If some off-site continuum exists and may be observed, it means that there needs to be some connection between off-site and observer's continuums. That's why some correspondence $ {\cal F} : { G} \stackrel{\cal F}{ \longleftrightarrow } \tilde { G} $ needs to be assigned between elements of two continuums $ {\it G} $: $ \{x: (x^0, \dots , x^n) \} $ and $ {\tilde G} $: $ \{\tilde{x}: (\tilde{x}^0, \dots , \tilde{x}^m) \} $. This transformation between continuums may be represented functionally (functions $f$ and $\tilde f$ may be of any kind) as: 

\numparts 
\begin{eqnarray}
\tilde{x}^j = f^j (x^0, \dots , x^n), \qquad & j = 0, \dots ,m \qquad &\mbox{or} \label{funcA} \\
x^i = \tilde{f}^i (\tilde{x}^0, \dots , \tilde{x}^m),  \qquad & i = 0, \dots , n \,. & \label{funcB}
\end{eqnarray}  
\endnumparts

If it is supposed that $ \tilde {\it G} $ has some metrics, so we may consider that some interval $d\tilde {s}^2 = \tilde {g}_{jl} d\tilde{x}^j d\tilde{x}^l $, where $ j,l=0..m $ is introduced. If we suppose that the off-site continuum $ \tilde {\it G} $ has its own invariant density $ \tilde {\mathfrak {R} } $, so we may write for him analogously to (\ref{INT}) the off-site Integral: 
$\tilde{\cal I} = \int_{\tilde{\it G}} {\tilde \mathfrak{R}} \, d^{m} {\tilde x} $. 

The functions $f$ and $\tilde f$ from (\ref{funcA}, \ref{funcB}) are very important for the ability to the observation of continuums. For example, if each element of two identical one-dimension continuums $G: x \in (-\infty .. \infty) $ and $\tilde {G} : \tilde {x} \in (-\infty .. \infty) $ is represented as a decimal number: $x =\dots a_3 \, a_2 \, a_1 \, a_0 , \, a_{-1}\, a_{-2} \, \ldots $; $\tilde x =\dots b_3 \, b_2 \, b_1 \, b_0 , \, b_{-1}\, b_{-2} \, \ldots $, and the correspondence $G \rightleftharpoons \tilde G$: $b_{2k+1}=a_{2k}, \quad b_{2k}=a_{2k+1}$, $k=0, \pm 1, \pm 2, \ldots$ is introduced, so, in spite of both $G$ and $\tilde G$ are continuous and measurable, and the correspondence is biunique, the ``mutual visible metrics'' can't be introduced. We understand as ``mutual visible metrics'' the metrics in one continuum, observable or visible from another one. 

If functions $f$ and $\tilde f$ from (\ref{funcA}, \ref{funcB}) are continuously differentiable, it gives an additional strong possibility for mathematical analysis. For example, on the field of definition, using $ d\tilde{x}^j = (\partial f^j /\partial x^i) dx^i = (\partial \tilde{x}^j /\partial x^i) dx^i $ (as usual, the summation on repeating indexes is meant), one can get the expression for the visible off-site metrics ($(ds^{\sim})^{2} \rightleftharpoons d\tilde{s}^{2} $): 

\begin{equation}
(ds^{\sim})^{2}=\tilde g_{jl}|_{\tilde x = f(x)} \case {\partial \tilde{x}^j }{\partial x^i} \case {\partial \tilde{x}^l }{\partial x^k} dx^i dx^k = g^\sim_{ik} dx^i dx^k; \quad i,k=0..n; \, j,l=0..m.
\label{metrics}
\end{equation} 

On the field of definition for the visible part of the off-site Integral one can get: 
$$ \tilde{\cal I} = \int_{\tilde{\it G}} {\tilde \mathfrak{R}} \, d^{m} {\tilde x} \rightleftharpoons \int_{\tilde{G}\cap G} {\mathfrak {R}}^{\sim} |\case {\partial \tilde{x}^j }{\partial x^i} | d^{n}x,\qquad  {\mathfrak {R}}^{\sim} = {\tilde \mathfrak{R}} |_{\tilde {x} =f(x)}, $$

where $| {\partial \tilde{x}^j }/{\partial x^i} |$ is a functional determinant. The case of $n \neq m$ needs a special analysis with the theory of implicit functions, but usually it means that the transformation between continuums is not uniquely defined. 

In an important particular case of the enclosed ($\hat G$: $\hat G \rightleftharpoons D \subset G$) and containing ($\check G$: $ G \rightleftharpoons \check D \subset \check G$) continuums, above correspondences may be represented by equalities: 

\numparts
\begin{eqnarray}
\hat{\cal I} = \int_{\hat{\it G}} {\hat \mathfrak{R}} \, d^{m} {\hat x} = \int_{D} {\mathfrak {R}}^{\wedge} |\case {\partial \hat{x}^j }{\partial x^i} | d^{n}x = {\cal I}^{\wedge},\qquad  {\mathfrak {R}}^{\wedge} = {\hat \mathfrak{R}} |_{\hat {x} = f(x)}; \label{INTA}\\
{\cal I} = \int_{{\it G}} {\mathfrak{R}} \, d^{n} {x} = \int_{\check D} {\mathfrak {R}}^{\vee} |\case {\partial {x}^i }{\partial \check{x}^j} | d^{m}{\check x} ={\cal I}^{\vee},\qquad  {\mathfrak {R}}^{\vee} = { \mathfrak{R}} |_{{x} = \check{f}(\check{x})}. \label{INTB} 
\end{eqnarray}
\endnumparts

In the general theory of relativity the invariant density $ {\mathfrak {R}} $ from (\ref{INT}) defines the space-time topometry and also depends on physical objects existing in this space-time. L.Landau and E.Lifshitz had described this fact in \cite {LL88} as: ``It is necessary, strictly speaking, to have a set of an infinite number of the bodies filling all space, like some ``medium". Such system of bodies together with connected to each of them arbitrarily clocks is a frame of reference in the general theory of relativity". Thus, the continuum $ {\it G} $ with invariant density $ \mathfrak {R} $ defines or generates the ``medium" -- the system including ``a set of infinite number of bodies" and corresponding conservation laws for these physical objects. 

The off-site continuum $ \tilde {\it G} $  may also generate its own system including its own physical objects, own space-time structure and conservation laws. This medium may differ from one generated by $ \it G $. We do not have reasons to deny the existence of such ``mediums". The argument, that if we do not observe something like that, so it does not exist and it is no sense to consider, is not convincing, because a question: what could we observe in this case? -- simply was never investigated. At least, the author does not know anything about such investigations. 

It is impossible to try to analyze all sets of possible continuums at once, but, certainly, not all of them may be physically realized. Therefore we shall enter minimally possible physical restrictions to make clear the physical background of the idea of the proposed hypothesis and to show the approach to the mathematical analysis of the off-site continuums.

\section{\bf Enclosed physical objects}
\label{sec:EPhO}

Further in paper we will consider the off-site continuums from ``our'' four-dimensional continuum $ {\it G} $: $ \{ x: (x^0, x^1, \dots, x^n) \} $, where $n=3$. We will denote the time as $x^0=\tau$, using the system of units, where speed of light $c=1$. From the definition of the enclosed continuum we have: $\hat G \rightleftharpoons D \subset G $. So, the observer may register the visual parameters of physical objects of the enclosed continuum only in some observable region $D \subset G$. We will consider that $D$ is a connectedness region in $G$. This way $D$ may be limited on some or on all four coordinates in $G$. In every case it will correspond to different types of visible off-site objects. 

The existence of stable off-site physical objects is the most important and principal question, so we will try to investigate this question first. ``Stable'' means unchangable in time ``in a whole'', i.e. in some principal characteristics at the observer's continuum. So, $D$ must be unlimited in time. We will consider the case of $D$ is limited in all space coordinates $x^\alpha,  \alpha=1..3 $. 

The visual attributes of off-site systems (off-site physical objects, conservation laws and topometry) will be perceived by the observer within the framework of his own continuum, hence, should satisfy to the physical laws generated by this continuum. We shall call this assumption as {\it a principle of compatibility} of off-site continuums. We will try to investigate the question: how may physical objects of the enclosed off-site continuum $\hat G$ percieve by the observer from ``our'' continuum? 

If some visible characteristic of the off-site enclosed object can be described in $G$ by some function, so one can determine this function on $D$ as $\Psi(x^i) \equiv \Psi(x^\alpha, \tau), i=0.. 3, \alpha=1..3 $ for $ \forall x^i \in {D} $ and $\Psi(x^i) \equiv 0$ for $ \forall x^i \not\in {D} $. To be observed the off-site physical objects should manifest themselves somehow, for example, with some well-known fields in a system of the observer. Let's assume, that the off-site object excites the wave type field $u(x^\alpha, \tau) $ and the function $\Psi(x^\alpha, \tau) $ is proportional to a source function $P(x^\alpha, \tau) $: $\Psi(x^\alpha, \tau) \sim P(x^\alpha, \tau) $, so, the excited field $u(x^\alpha, \tau) $ needs to satisfy to the wave equation: 

\begin{equation}
\left(\frac{1}{V^2}\frac{\partial^2 }{\partial {\tau^2}} - \bigtriangledown ^2 \right) u(x^\alpha,\tau) = P(x^\alpha,\tau), \qquad \bigtriangledown ^2 = \sum_\alpha \frac{\partial^2 }{\partial {x^{\alpha}}^2}. 	
\label{waveeq}
\end{equation}

According to a principle of compatibility declared above, the solutions can't contadict to conservation laws in the system of the observer. It means, for example, that $P(x^\alpha, \tau) $ cannot be an infinite energy source in the system of the observer. It is possible only if the excited fields $u(x^\alpha, \tau) $ do not transfer energy or if they are localized in $G$. It is possible to satisfy to the condition of the field localization only for the specific source functions $P(x^\alpha, \tau) $. So, we need to find such source functions $P_{st}(x^\alpha, \tau) $, that the fields $u_{st}(x^\alpha, \tau) $ excited by them according to the wave equation \eref{waveeq} are localized in $G$. It is possible only if the excited fields compensate each other outside the region $D \subset G$. 

\begin{figure}[ht]
	\label{fig:Excite}
	\begin{center}
	\includegraphics*[height=100mm, angle=270]{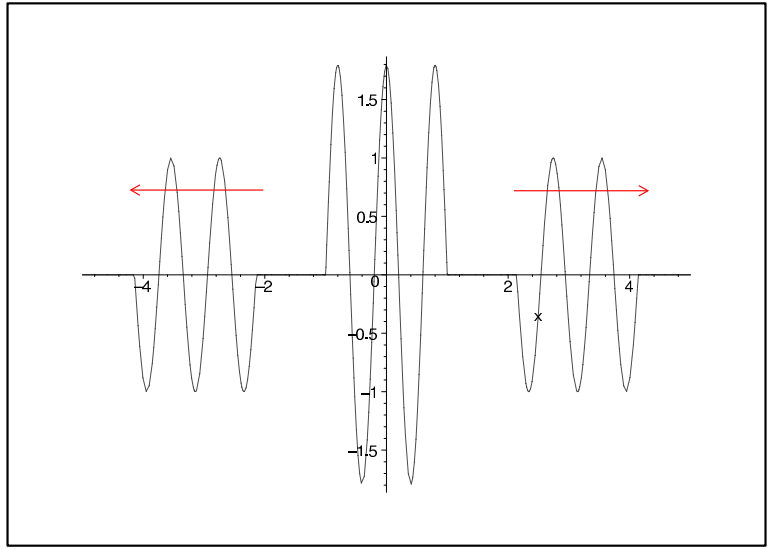}
	\caption{Excited fields of the enclosed physical object.}
	\end{center}
\end{figure}

The principle of such compensation may be easily illustrated in one-dimensional case by the following example. The generalized function $P(x, \tau) = \left[\delta (x+a) + \delta (x-a) \right] \Theta(\tau) \sin(\Omega \tau)$, where $\delta$ is a $\delta$-function, $ \Theta(\tau) =0$ at $\tau <0$ and $\Theta(\tau) =1$ at $\tau \geq 0$, describes two identical synchronous sources in space-points $x=-a$ and $x=a$. The generalized functions are described, for example, in \cite{Vlad81}. They usually allows to find the fundamental solutions and to reveal the basic properties of the general solutions. At a time-moment $\tau=0$ the sources start to excite $\sin$-like waves with the frequency $\Omega$: $\sin[\Omega (\tau - |x-a|) / V]$ and $\sin[\Omega (\tau - |x+a|) / V]$. Waves spread out the sources with the speed $V$ along and opposite the $X$-axis. If the wave from one source reaches the another source in the antiphase: $2 \Omega a /V =-\pi + 2 \pi n $, waves will compensate each other. So, one will have $ \forall  \tau : u(x,\tau) = 0$ for $ |x|> a $, while there exists the standing wave between two sources: $u(x,\tau)\not\equiv 0$ for $ |x| <a $. As far as it needs time $2a/V$ for the wave to reach from one source to another, during this time period the waves won't be compensated. So, at the beginning, the observer will also register two wave trains or quanta flowing out from the rest stable source system. This process is illustrated on Figure~\ref{fig:Excite} for $a=1$. Generally, the parameters of the flowing out wave trains depend on the synchronization of the sources. For example, if one source start to excite waves through time $2a/V$ after another source, it would be only one flowing out object, but of double length. 

Thus, excited fields may be localized in some limited space region in the system of the observer. The correlation between space sizes and source frequency is needed for the existence of the stable off-site physical object. We will call such conditions as {\it conditions of quantization}. It is a consequence only of the energy conservation law at the observer's continuum. So, the stable off-site object may be observed only in stable states. The ``birth'' of the off-site object or its changing from one stable state to another will always be accompanied by emitting or absorbing the quanta of waves, which may be called as ``parts of field'' or field particles. 

If only stable states of the off-site objects are interested in, so they may be found by considering the Helmholtz equation, corresponding to \eref{waveeq}, since the specific source distributions correspond to every source frequency. The problem may be set as follows: to find such $P^\Omega (x^\alpha)$ at which corresponding solutions $U^\Omega(x^\alpha) $ of the Helmholtz equation are localized inside some closed connectedness region $ D_0 \subset G$:

\begin{equation}
\fl \left( \nabla ^2 + \frac{\Omega^2}{V^2} \right) U^\Omega(x^\alpha) = - P^\Omega (x^\alpha), \qquad U^\Omega(x^\alpha) = \cases {\not\equiv 0 & for $x^\alpha \in D_0$, \\ = 0 & for $x^\alpha \not\in D_0$, }. 
\label{Helm}
\end{equation}

Here $P^\Omega (x^\alpha)$ is a distribution of the complex amplitudes of wave sources with frequency $\Omega$, $U^\Omega(x^\alpha) $ is a distribution of the complex amplitudes of excited fields. The additional conditions for $U^\Omega(x^\alpha) $ on the closed region $ D_0 \subset G$ defines the connection between space distribution and the source frequency. It is exactly these conditions we have called as {\it conditions of quantization}. 

For example, in the last example of the one-dimensional field excitation, the source function may be written as $P^\Omega(x) = \delta(x+a) + \delta(x-a) $ and one can get the stable state solutions: 

$$
U^\Omega(x) = \cases { \case{ i V}{\Omega} e^{- i\Omega a/V} \cos(\Omega x/V) & for $|x | <a$, \\  
\case{ i V}{\Omega} e^{- i\Omega x/V} \cos(\Omega a/V) & for  $|x |> a$.  } 
$$

At $ \Omega a/V =-\pi/2 + \pi n $: $U^\Omega (x) = 0$ for $ |x |> a $, while  $U^\Omega(x) \not\equiv 0$ for $ |x | <a $. 

In a 3D-case the solutions for stable states also exist. For example, for spherically symmetric source function $ P^\Omega (x^\alpha) = \delta(r-a) /(4 \pi a^2)$,  $ r =\sqrt{x_\alpha x^\alpha} $ one can get: 

$$U^\Omega(r) = \cases{ \case{- i V}{\Omega ar} e^{-i\Omega a/V} \cos (\Omega r/V) & for $0< r  <a$, \\ 
\case{ - i V}{\Omega ar}  e^{ -i \Omega r/V} \cos (\Omega a/V) & for $r >a$.}
$$

If $ \Omega a/V = - \pi/2 + \pi n $ fields appear located inside the sphere of the radius $ a $. 

Thus, the solutions of the wave equation \eref{waveeq} for stable states may be represented in complex amplitudes as: 

\begin{equation}
 P(x^{\alpha},\tau) \doteqdot \sum_{\Omega}P^\Omega (x^{\alpha}) e^{i\Omega \tau} , \qquad 
 u(x^{\alpha},\tau)\doteqdot \sum_{\Omega}e^{i\Omega \tau} \sum_k U^\Omega _k (x^{\alpha}).  
\label{wfunc}
\end{equation}

At the beginning of this Section we have introduced the function $\Psi(x^{\alpha},\tau)$ as a function corresponding to some {\it visible} characteristic of off-site physical object. We have also assumed that $\Psi(x^\alpha, \tau) \sim P(x^\alpha, \tau) $. But it is exactly $u(x^{\alpha},\tau)$ a visible (from $G$) characteristic of the off-site object. If identifying $\Psi(x^{\alpha},\tau)$ with $u(x^{\alpha},\tau)$ and applying $\Psi(x^\alpha, \tau) = - \mu^2 P(x^\alpha, \tau) $, one can immediately get the Klein-Gordon equation for free particle: 

$$ 
\left(\frac{1}{V^2}\frac{\partial^2 }{\partial {\tau^2}} - \bigtriangledown ^2 \right) \Psi(x^\alpha,\tau) = - \mu^2 \Psi(x^\alpha,\tau).	
$$

The supposition $\Psi(x^\alpha, \tau) = - \mu^2 P(x^\alpha, \tau) $ also seems quite logical, because, as we have seen before, the source function needs to be in antiphase ($e^{i \pi}=-1$) to the spreading wave to compensate the external fields. The identification $\Psi(x^{\alpha},\tau)$ with $u(x^{\alpha},\tau)$ and $P(x^\alpha, \tau) $ lets the expressions \eref{wfunc} to be appropriate for $\Psi(x^{\alpha},\tau)$. This way, $\Psi(x^{\alpha},\tau)$ has a meaning of the wave function of the physical object. 

We have presented these identifications to make clear the background of the Klein-Gordon equation and the wave function from our approach point of view. Of course, though these suppositions seem quite logical, they reduce considerably the set of possible solutions of the equation \eref{waveeq}. Moreover, generally, the function $\Psi(x^{\alpha},\tau)$ is not identical to $u(x^{\alpha},\tau)$. Anyway, the Klein-Gordon equation may be considered as a particular case of our set of the problem to the equations \eref{waveeq} and \eref{Helm}. 

The electromagnetic (em) waves are appropriate for the waves excited by the off-site objects. The analysis of the em waves excitation is more complicated, because it needs to analyse the system of equations like \eref{waveeq} for each corresponding vector component. But, in principal, the analysis doesn't differ from the considered above. By means of matrix factorization of the Klein-Gordon equation \cite{Vlad81}, introducing and considering the spinor fields, one may come to the Dirac equation. From this point of view, the Dirac equation may also be considered as a particular case of our approach. 

The process of the em wave excitation inside some closed space region is investigated quite well in the accelerator physics, where the resonant excitation of the cavities are described. The localization of the em fields in the cavity due to electric or magnetic walls may be represented as the presence of some off-site sources compensated the outer fields. It almost looks like our model of stable off-site object generating by the surface sources. 

It is important for us that the compact self-consistent em fields $u(x^{\alpha},\tau)$ of the enclosed object will possess explicit quantum properties for the observer, that is a consequence of the principle of compatibility of off-site objects and, in particular, of the energy conservation law in the system of the observer. These fields are good candidates for a role of weak interactions. Such interpretation is already confirmed experimentally by a fact of equivalence of weak and electromagnetic interactions at high energy levels.  

If the region $D$: $\hat G \rightleftharpoons D \subset G$ are limited on both all space and time coordinates, the observer should see the occurrence and disappearance in the limited area of space of some {\it virtual} objects. The source function of such objects may be written as $P(x^\alpha,\tau)= P_0(x^\alpha,\tau) \Theta(\tau-T_0) \Theta(T_1-\tau)$, where $\left(T_0 , T_1 \right)$ is the time interval of existence of the enclosed object in the system of the observer. The occurrence and disappearance of the off-site physical object will be accompanied by the emission and absorption of quanta of the field excited by objects in the system of the observer. Most likely, the continuous process of quanta exchanging should be initiated inside some system of the off-site virtual objects. Generally speaking, this process does not contradict to modern representations of the internal structure of the physical vacuum. 

Detailed study of every possible variants of the perception of the enclosed off-site objects in system of the observer is very interesting and extended task. It may be a subject for the further detailed researches. 

Note, that in this section we haven't made any suppositions about the off-site continuums by themselves. We have only demanded the non-contradictory observation of the enclosed objects in the observer's continuum. 

\section{\bf Visible structure of off-site continuums}
\label{sec:Struct} 

To investigate in details the visible structure or topometry of continuums, we need to suppose that, at least, off-site continuums have such structure or topometry by themselves. So, we consider that some metrics $d\tilde{s}^2=\tilde{g}_{jl}d\tilde{x}^j d\tilde{x}^l$, $j,l = 0, \dots , m$ is defined in the off-site continuum $\tilde G$. Another supposition concerns to the possibility of the observation of this structure from the continuum of the observer. In the example to (\ref{funcA},\ref{funcB}) it was shown, that some transformations can make metrics undefinable, even if it exists. That's why we should demand from functions $ f $ and $\tilde f$ from (\ref{funcA}, \ref{funcB}) to be continuously differentiable on the field of definition. So, we're coming straight to the analysis of the {\it visible off-site metric tensor} from \eref{metrics}:
$g^\sim_{ik} = \tilde g_{jl}|_{\tilde x = f(x)} \case {\partial \tilde{x}^j }{\partial x^i} \case {\partial \tilde{x}^l }{\partial x^k}; \quad i,k=0..n; \, j,l=0..m .$

L.Landau and E.Lifshitz \cite {LL88} have given the following physical interpretation of parameters of ``usual'' metric tensor in the system of the observer: ``It is necessary to emphasize a difference between meanings of a condition $g_ {00}> 0 $ and a condition of the certain signature (signs on principal values) of the metric tensor $g_{ik} $. The tensor $g_{ik} $ non-satisfying to the second one of these conditions, can't correspond to any real gravitational field at all, i.e. the metrics of the real space-time. Non-fulfillment of the condition $g_{00}> 0 $ would mean only, that the corresponding frame of references can't be realized by real bodies; thus if the condition on principal values is carried out, it is possible to achieve to $g_{00} $ becomes positive by appropriate transformation of coordinates". 

\begin{figure}[ht]
	\begin{center}
	\includegraphics*[width=50mm, angle=270]{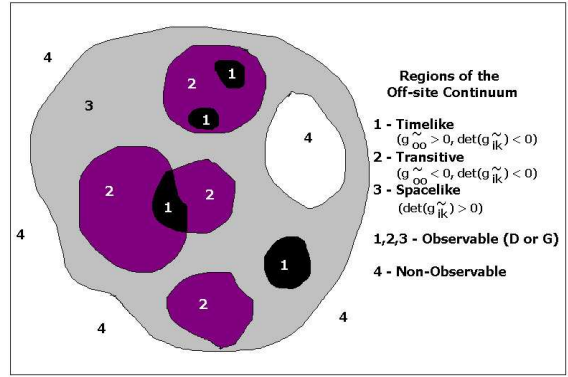}
	\caption{The structure of the off-site continuums}
	\label{fig:Struct}
	\end{center}
\end{figure}

Of course, the visible off-site metric tensor $ g^\sim_{ik} $ is not at all a tensor in $G$, but we will consider that its parameters determine the observable physical space-time structure of $ {\tilde G} $ and the visual properties of the off-site physical objects. Following by L.Landau and E.Lifshitz \cite {LL88} we may separate conditionally the off-site continuum into three areas: {\it the timelike region} ($ g^\sim_{00} > 0 $ and $ \det(g^\sim_{ik}) < 0 $); {\it the spacelike region} ($ \det ( g^\sim_{ik}) > 0 $) and {\it the transitive region} covering the rest part ($ g^\sim_{00} < 0 $ and $ \det(g^\sim_{ik}) < 0 $). The schematic drawing of the structure of the enclosed continuum inside the observation region $D $ is presented on Figure~\ref {fig:Struct}. 

The observable time and distance may be defined only in timelike regions. The observable off-site time can be determined from (\ref{metrics}) assuming $dx^{\alpha} =0, \alpha=1,2,3 $: $d\tau^{\sim} = \sqrt {g^{\sim}_{00}} dx^0 $, the observable distance: $ (dl^{\sim})^2 =\gamma^{\sim}_{\alpha \beta} dx^{\alpha} dx^{\beta},\, \gamma^{\sim}_{\alpha \beta} = - g^{\sim}_{\alpha \beta} + g^{\sim}_{0 \alpha} g^{\sim}_{0 \beta}/g^{\sim}_{00} $. So, the ``real space-time" and ``real bodies" may be observed only in timelike regions. Therefore, if the off-site physical objects exist in timelike regions, they can be detected locally in the system of the observer. It means, for example, that they may take part in inelastic scattering with other particles. 

The observation of off-site physical objects from spacelike regions may be extremely unusual because of the discrepancy between metrics. The off-site physical objects in $\tilde G$ and their perception in $G$ may not coincide completely with each other. It may be a lot of different possibilities. For example, due to the absence of the space correspondence, they will be observed as something amorphous, distributed in space. Even, in some particular cases, it may seem, that cause-effect chains are broken in off-site continuums. For example, one object may exist simultaneously in quite different regions of the observer continuum, etc. 

There may be a mixed perception of objects in transitive region depending on the concrete reference frame of the observer. It means that it is possibile to observe the ``real parameters'' of the off-site objects from one reference frame of the observer, but it won't coincide with the perception from another frame of references. 
 
The general picture is even much more complicated, because the components of the visible metric tensor depend on all coordinates $x^i $, including $x^0 = \tau$, in the system of the observer. So, the internal structure of the enclosed continuum may be dynamical for the off-site observer. Note, that according to analysis of Section~\ref{sec:EPhO}, there need to be some restrictions on possible visible internal structures of the enclosed continuums. 

Physically, from the observer's point of view, the enclosed objects may have some ``hard'' parts in timelike regions surrounded by something amorphous. These ``hard'' parts would be observed as captured inside the space region $D$. Such capture or confinement should seem as the extremely unusual phenomenon. At the standard field interpretation, it is equivalent to the presence of some forces holding these objects inside the space area $D$. It should seem, that the quantity of these forces grows considerably with the distance or with approaching to the borders of the region $D$. Analogies with quarks, quakr-gluon plasma, the phenomenon of confinement and strong interactions are arising at once. 

The analysis of the structure of containing continuum and its physical objects does not differ considerably from just considered enclosed continuums. The observable region for $\check G$ will coinside with the whole continuum $G$. Hence, the whole continuum $G$: ${G} \rightleftharpoons \check D \subset {\check G} $ may be separated on three regions: timelike, spacelike and transitive, with the corresponding properties identical to ones just considered. 

The only thing, that the observer from $G$, generally speaking, can not even perceive the boundedness of his own continuum on spatial and time coordinates in an off-site containing continuum $ {\check G}$. So, the essential distinction is that there exists the fourth {\it non-observable area} for the observer from $G$ in the containing continuum $\check G$ (see \Fref{fig:Struct}). The non-observable area is outside the observer's continuum $G$, so, in principal, this region is inaccessible for the observer. However, the influences from physical objects of this area may be registered by the observer, at least, because they should change the space-time structure of $G \rightleftharpoons \check D$. It means that the observer may see the metric deformation of his own continuum without visible reasons, and, hence, may detect the changing on seen parameters and movements of physical objects in this area. It is the most probable to detect such deformations on macro-scale, at vast distances. In the GTR, the energy is responsible for metrics deformation, so the objects from non-observable areas will be percieved as some invisible, dark energy. There are good candidates for this role of such invisible off-site physical objects of off-site continuums in astrophysics. It is the dark energy discovered not so long ago. 

Let's continue analogies. The off-site physical objects of containing continuum from spacelike and transitive regions may look more ``material'' for the observer in comparison with the dark energy, similar to the dark matter. Some galaxies or star systems may correspond to the off-site physical objects of containing continuum from the timelike regions. The visible metrics of such regions may be quite different and may differ considerably from the standard model, predicted by general theory of relativity, because this is a visual metrics of another continuum. By the way, such visual metrics from different off-site continuums may differ from each other, depending on the type of connection (the type of transformation) with the continuum of the observer. 

The streams of particles emitted by physical objects of non-observable area also may be detected. These particles can penetrate the area $\check D$, that can be perceived as the streams of particles originated by an invisible source, from the ``empty" space of the observer's continuum. Corresponding reports about the registration of intensive streams of particles from ``empty" areas of the Universe have also appeared not long time ago.  

In section~\ref{sec:EPhO} there were investigated the fields in the containing continuum induced by objects of the enclosed continuum. These fields exist in $G $ and also can be registered by the observer, for example, as the streams of particles penetrating $G$ in all directions. These particles, generally speaking, differs from just mentioned above, because they are excited by the physical objects of the observer's continuum, but not by the invisible source. So, they should correlate with the sizes (or the ``age" in the standard model of the Universe) of our continuum $G$. The relic radiation is a good candidate for this role. 

It's necessary to tell a few words about an opportunity of the existence and observation of off-site objects in some ``average", ``human" scale, the scale of the observer. It is possible to assume, that the power density of containing system defines the scale of observation of off-site enclosed objects. In other words, the existence of the high energetic off-site objects in micro-scale interferes with their stable existence at the ``average" scale. 

It would be quite interesting and logical to use expressions (\ref{INTA}, \ref{INTB}) for further analysis. It is possible to guess that the invariant Integral of off-site continuum certainly needs to have some physical meaning in the system of the observer. Analogies are just arising. However, such analysis would lead to the additional physical suppositions, that needs the accurate detailed researches. It was already mentioned before, that it is not at all the only way to continue investigations of off-site continuums. 

\section*{\bf Conclusion}

The hypothesis of the feasible existence of off-site continuums was investigated in this paper. 

It was found that in a scale of the elementary particles, the visual properties of some off-site physical objects correspond to the similar properties of quantum-mechanical and quantum-physical objects. Some of them satisfy to the Klein-Gordon and Dirac equations, and described by the functions similar to the wavefunctions. Quantization is an essential observable property of stable off-site physical objects and is a consequence of the energy conservation law in a continuum of the observer. There were found the obvious correlations with weak and strong interactions, quarks and the phenomenon of confinement. The modern model of physical vacuum has also obvious correlations with some types of the off-site physical objects. The suggested approach considerably expands opportunities of search and the analysis of new physical objects of a microcosm.

In a scale of astrophysical objects, the appearance of off-site objects is quite similar to the influence of the ``dark matter" and the ``dark energy" discovered few years ago. Analogies to known relic radiation are also looked through. Such point of view may appear useful at the investigation of actual principal problems of modern astrophysics. 

The proposed hypothesis of the existence of off-site continuums may permit to unify the relativistic and quantum-physical approaches into the non-contradicting system. It may be expected that the analysis of the off-site physical objects could help to discover a lot of new physical objects in our reality and to investigate their quite unusual properties both in micro- and macro-scales.

%\label{lastpage}
\end{document}